\documentclass[prl, superscriptaddress, showpacs]{revtex4-1}
\usepackage{amsmath}
\usepackage{amssymb}
\usepackage{bm}
\usepackage{amsfonts}
\usepackage{amsthm}
\usepackage{graphicx}

\newcommand{\bk}{{\boldsymbol k}}

\newcommand{\bq}{{\boldsymbol q}}
\newcommand{\bp}{{\boldsymbol p}}

\newcommand{\fbf}{{\boldsymbol f}}
\newcommand{\bF}{{\boldsymbol F}}

\newcommand{\bv}{{\boldsymbol v}}

\newcommand{\bx}{{\boldsymbol x}}
\newcommand{\by}{{\boldsymbol y}}
\newcommand{\bz}{{\boldsymbol z}}

\newcommand{\bX}{{\boldsymbol X}}

\newcommand{\bxi}{{\boldsymbol \xi}}

\renewcommand{\div}{\operatorname{div}}

\begin{document}
\title{
Kinetic equation for spatially averaged molecular dynamics}
\author{Alexander Panchenko}
\author{Kevin Cooper}
\author{Andrei Kouznetsov}
\affiliation{Department of Mathematics, Washington State University}
\author{Lyudmyla L. Barannyk}
\affiliation{Department of Mathematics, University of Idaho}
\begin{abstract}
We obtain a kinetic description of spatially averaged dynamics of particle systems.
Spatial averaging is one of the three types of averaging relevant within the Irwing-Kirkwood procedure (IKP), a general method for deriving macroscopic equations from molecular models. The other two types, ensemble averaging and time averaging, have been extensively studied, while spatial averaging is relatively less understood. We show that the average density, linear momentum, and kinetic energy used in IKP can be obtained from a single average quantity, called the generating function. A kinetic equation for the generating function is obtained and tested numerically on Lennard-Jones oscillator chains.  
\end{abstract}
\maketitle
In 1950, Irwing and Kirkwood \cite{Kirkwood} proposed an averaging method for deriving macroscopic theories from molecular description. Three types of averages are relevant
within the Irwing-Kirkwood procedure (IKP):
ensemble averages \cite{Kirkwood, Noll}, time averages \cite{mb}, and space averages \cite{Hardy, PBG, TPF}. All three types can be used either separately or together. Ensemble averaging and time averaging are well understood, while spatial averaging is relatively less explored. Studying space averaging is useful because (i) space-time averages represent the most realistic model of macroscopic measurements in a single experiment; (ii) the number of repetitions in engineering experiments can be too small for accurate sampling of the underlying probability distribution;
(iii) spatial averages are easy to compute in molecular dynamics (MD) simulations whereas ensemble averaging requires costly integrations in phase space, and long-time averages may be inaccessible because of time-scale limitations of MD algorithms.

Thus it makes sense to ask the following question. What information can be obtained from spatial averages of a single MD run? 
The purpose of this Letter is to derive a kinetic description of spatially averaged molecular dynamics. We are particularly interested in dense fluids and soft matter. It is well known  \cite{boon-yip, evans, dorfman-cohen65, dorfman-cohen72} that extending classical kinetic theory \cite{Resibois} to such systems is difficult due to non-analytic behavior of expansions with respect to density. 

Consider a system of $N$ classical particles of equal mass $m=M/N$ confined to the domain $\Omega$, and interacting with short-range forces generated by a pair potential $U$.  Microscopic state variables are positions $\bq_j(t)$ and momenta $\bp_j(t)$. Mesoscopic behavior of the system can be characterized using 
spatially averaged density $\overline{\rho}$, linear momentum $\overline{\bp}$, and kinetic energy $K$
\begin{eqnarray}
\label{standard-averages}
\overline{\rho}(t, \bx) & = &\sum_{j=1}^N m \psi_\eta(\bx-\bq_j(t)),\nonumber\\
\overline{\bp}(t, \bx) & = & \sum_{j=1}^N \bp_j(t)\psi_\eta(\bx-\bq_j(t)),\\
K & = & \frac{1}{2m}\sum_{j=1}^N \bp_j(t)\cdot\bp_j(t) \psi_\eta(\bx-\bq_j(t)).  \nonumber
\end{eqnarray}
The mesoscopic length scale $\eta$  is much larger than the characteristic interparticle distance, but may be much smaller than the extent of the whole system.  For each $\eta$, the averages in (\ref{standard-averages}) satisfy exact continuum-style balance equations of mass, momentum, and 
energy \cite{Hardy, mb}. 

The window function $\psi$ is normalized by requiring $\int\psi(\bx) d\bx=1$,  and then scaled by $\eta$ so that $\psi_\eta(\bx)=\eta^{-d} \psi(\bx /\eta)$, where $d$ is the physical space dimension. Scaling ensures that $\psi_\eta(\bx-\bq_j)$ converges to $\delta(\bx-\bq_j)$ in the limit $\eta\to 0$. In this limit one recovers the phase-space densities similar to the densities used in  IKP \cite{Kirkwood} and statistical hydrodynamics \cite{boon-yip}. 
Increasing  $\eta$ increases the number of particles within the averaging volume (the support of $\psi_\eta$). This has the effect of filtering out high frequency oscillations. For example, the Fourier transform ${\mathcal F}_{\bx\to\bk} \overline{\rho}$ 
is equal to $\widehat\psi(\eta\bk) \sum_{j=1}^N m e^{i \bk\cdot \bq_j}$, where $\widehat\psi$ is the Fourier transform of $\psi$, and  $\sum_{j=1}^N m e^{i \bk\cdot \bq_j}$ is the Fourier transform of $\sum_{j=1}^N m \delta(\bx-\bq_j)$. For larger $\eta$, the filter function $\widehat\psi(\eta\bk)$ is more localized near $\bk=0$ \cite{BP12}.

To derive a kinetic equation, we introduce another spatial average, called the {\it generating function}
\begin{equation}
\label{gen0}
Q(t, \bx, \bz)=\sum_{j=1}^N m e^{im^{-1}\bp_j\cdot \bz} \psi_\eta(\bx-\bq_j).
\end{equation}
As noted above, for larger $\eta$ the Fourier transform ${\mathcal F}_{\bx\to \bk}$ of $Q$ becomes progressively more localized near $\bk=0$. Therefore, $Q$ is expected to be slowly varying provided the averaging scale is sufficiently large.

The continuum averages $\overline{\rho}, \overline{\bp}$ and $K$ can be obtained from $Q$ 
as follows: $\overline{\rho} = Q(t, \bx, 0)$, $\overline{\bp} =  -i\nabla_{\bz} Q(t, \bx, 0)$, $ K = -\frac 12\nabla_\bz\cdot\nabla_\bz Q(t, \bx, 0)$.  These equations can be related to the standard kinetic theory moment expressions by noting that the Fourier transform ${\mathcal F}_{\bz\to \bxi}$ of $Q$ is the spatially coarsened phase space density
$$
f(\bx, \bxi)=\sum_{j=1}^N m \delta(\bxi-m^{-1}\bp_j) \psi_\eta(\bx-\bq_j)
$$
that is similar to the density used by Mori \cite{mori72}.  Therefore, $\bz$-differentiation of $Q$ corresponds to multiplication of $f$ by 
$i\bxi$, and evaluation at $\bz=0$ corresponds to integration of moments of $f$ with respect to $\bxi$. Using $Q$ instead of $f$ may be more convenient because, instead of integrating $f$ over all $\bxi$, averages of interest can be now computed by differentiating 
$Q$ and then evaluating at $\bz=0$. To accomplish this, one only needs to know $Q$ for $\bz$ in a neighborhood of zero.

Taking time derivative in Eq. (\ref{gen0}) and using Newton's equations $\dot\bp_j=\fbf_j$ yields the exact evolution equation
\begin{equation}
\label{eq:Q-exact}
\partial_t Q- i \nabla_\bx \cdot \nabla_\bz Q_\eta
 =i \bz \cdot {\bF},
 \end{equation}
 where
\begin{equation}
\label{eq:f-exact}
\bF(t;\bx, \bz) =
	\sum_{j=1}^N \fbf_j e^{i m^{-1}\bp_j(t) \cdot \bz} \psi_\eta(\bx - \bq_j(t)),
\end{equation}
and $\fbf_j$ is the total force acting on a particle $j$. This equation is not coarse-grained, since one must know all $\bq_j(t)$ and $\bp_j(t)$ to evaluate $\fbf_j$. 

The principal contribution of this work is the closed form approximation of the exact $\bF$ in (\ref{eq:f-exact}) by an operator acting on $Q$. The first step in the derivation is to approximate $Q$ and $\bF$ by integrals. In doing so, we deviate from the standard phase space description of dynamics and think instead of a physical domain 
containing moving particles.  From this point of view, the natural objects are micro-scale continuum deformation $\tilde\bq(t, \bX)$ and velocity $\tilde\bv(t, \tilde\bq)$ in the physical space-time \cite{PBG, PBC}. At each $t$, these fields interpolate, respectively, particle positions and velocities. Using the interpolants to approximate sums by integrals, we find
\begin{eqnarray}
\label{Q-int-ap}
Q & \approx  &
\frac{M}{\mathcal{V}_\Omega}\int_{\Omega} e^{i\tilde{\bv}(t, \tilde{\bq}(t, \bX))\cdot \bz} \psi_\eta(\bx-\tilde\bq(t, \bX)) d\bX\\
& = &
\frac{M}{\mathcal{V}_\Omega}\int_{\Omega} e^{i\tilde{\bv}(t, \by)\cdot \bz} \psi_\eta(\bx-\by)J(t, \by) d\by,\nonumber
\end{eqnarray}
where $\mathcal{V}_\Omega$ is the volume of $\Omega$, and $J=\left|\mathrm{det}\nabla\tilde\bq^{-1}\right|$.
Similarly,
\begin{eqnarray}
\label{f-int-ap}
\hspace*{0.7cm}{\bF} & \approx &
\frac{N}{{\mathcal{V}_\Omega}}\int_{\Omega} e^{i \bz\cdot \tilde{\bv}(t, \tilde\bq(t, \bX))}\psi_\eta(\bx -\tilde{\bq}(t, \bX))
\sum_{k=1}^N -U^\prime\left( \tilde{\bq}(t, \bX)-\bq_k(t)\right)d\bX \nonumber \\
& = &
\frac{N}{{\mathcal{V}_\Omega}}\int_{\Omega} e^{i \bz\cdot \tilde{\bv}(t, \by)}\psi_\eta(\bx -\by)
\sum_{k=1}^N -U^\prime\left( \by-\bq_k(t) \right)J(t, \by)d\by.
\end{eqnarray}

The fine scale quantities in (\ref{f-int-ap}) are $\tilde{\bv}, J$, and positions $\bq_k$. They have to be approximated in terms of the averaged quantity $Q$. Since $\psi_\eta$ is close to the delta-function for small $\eta$,  one can write
$e^{i\tilde{\bv}(t, \by)\cdot \bz}J(t, \by) \approx \frac{\mathcal{V}_\Omega}
{M} Q(t, \by, \bz)$.  To approximate $\bq_k$, we use the average density $\overline{\rho}(t, \bx)=Q(t, \bx, 0)$ and replace 
$\bq_k(t)$ with $\widehat{\bq}_k(t)$ obtained by (i) splitting the physical domain into mesoscopic cells, and (ii) placing particles periodically inside each cell so that the average density of this packing is equal to $Q(t, \bx, 0)$ at the center of the cell. We note that the periodic placement can be replaced by a random placement sampled from an appropriate distribution. The number of particles placed inside each cell should be still consistent with the measured value of $Q(t, \bx, 0)$ inside that cell. Combining equations, we obtain the closed form approximation
\begin{equation}
\label{closed-form}
\bF(t, \bx, \bz)\approx \overline{\bF}(t, \bx, \bz)=\frac{N}{M}\int Q(t,\by, \bz) \psi_\eta(\bx-\by) \sum_{k=1}^N -U^\prime\left( \by-\widehat \bq_k(t) \right) d\by
\end{equation}
and the corresponding kinetic equation
\begin{equation}
\label{kinetic-final}
\partial_t Q- i \div_\bx \nabla_\bz Q_\eta
 =i \bz \cdot \overline{\bF}.
\end{equation}

To test the closure formula (\ref{closed-form}), we ran an MD simulation
of a 1-D system with $N=10,000$ particles, interacting with the Lennard-Jones 
potential $U$. 
The potential was truncated 
at the distance $20/N$. We took
${\mathcal V}_\Omega = 1$ and $M=1$. Periodic boundary conditions were imposed, and the equations of motion were solved using
the Verlet algorithm.  We considered two sets of initial conditions. In the first set, the initial positions were 
uniformly spaced, and the initial velocities were prescribed using a centered, piecewise polynomial, approximate Gaussian pulse for the middle third of the particles.  In the second set,  the initial velocities were set to zero, and the initial positions were chosen to vary sinusoidally.

The averages were generated using the window function
$\psi(x)$ eequal to 
$\frac{15}{16}(1+x)^2(1-x)^2$ when $\vert x\vert < 1$, and equal to zero otherwise.
We computed $Q$ and $\bF$ using $\bq_j, \bv_j$ from the MD simulation, and then used the obtained $Q$ to evaluate $\overline{\bF}$ from (\ref{closed-form}). The integral quadrature in (\ref{closed-form}) was implemented by first generating a uniform grid with step size $h$, and then scaling the grid points by the density, so that
$\sum_{k=1}^N -U^\prime
\left(
y_i-\hat{q}_k(t)
\right)
\approx
\sum_{k=1\atop k\ne i}^N -U^\prime
\left(
\frac{h(i-k)}{Q(t,y_i,0)}
\right).
$


\begin{figure}[h]\centerline{
\includegraphics[width=.45\textwidth]{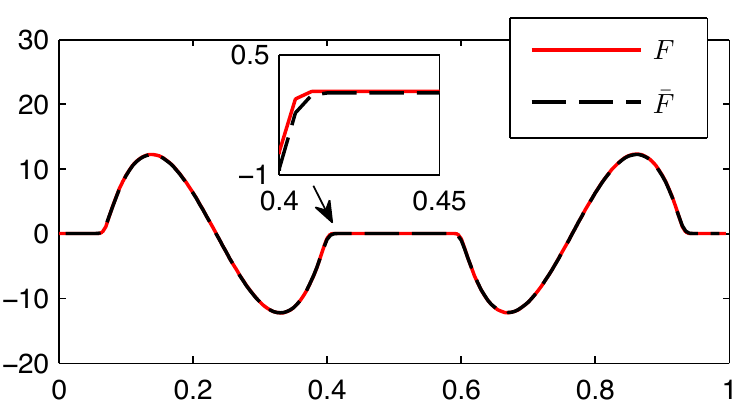}
\hspace{20pt}
\includegraphics[width=.45\textwidth]{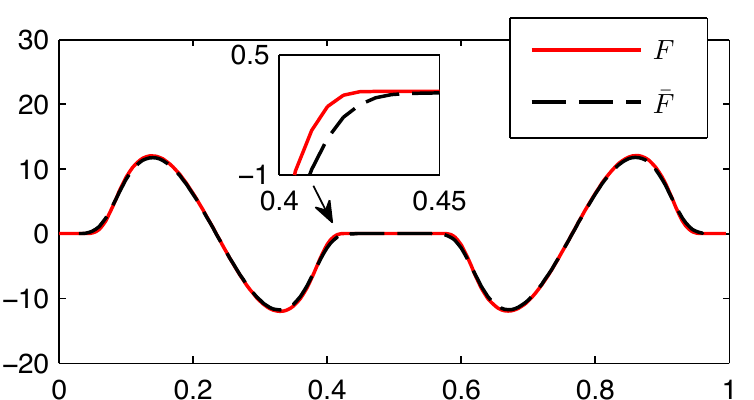}
}
\caption{\label{fig:f2} 
Computed  flux and closure approximation for $N=10^4$, $t=0.002$, $z=0.2$,  $\eta=0.01$ (left panel) and  $\eta=0.03$ (right panel).}
\end{figure}


\begin{figure}[h]
\centerline{
\includegraphics[width=.4\textwidth]{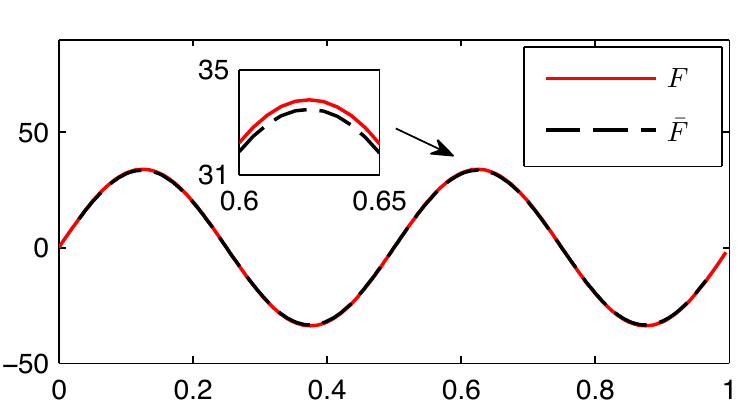}
}
\caption{\label{fig:f4}
 Computed  flux and closure approximation for $N=10^4$, $t=0.002$, $\eta=0.03$, and $z=0$.
}
\end{figure}

%


The approximation of the flux $\bF$ by the closed-form $\overline{\bF}$ is quite accurate
over the range of parameters we used.  Figure \ref{fig:f2} shows the real part of $\bF$
computed from Eq. (\ref{eq:f-exact}),  compared with the meso-scale flux $\overline{\bF}$ computed from the closure 
Eq. (\ref{closed-form}).  Increasing $\eta$ has only a slight effect on the quality of the approximation, as shown in the right panel of Figure \ref{fig:f2}. Approximately the same accuracy was obtained for all $z$ with $|z|\leq 0.2$. An example with $z=0$ is shown in Figure
\ref{fig:f4}. In this range of $z$, and for $\eta$ values between $0.01$ and $0.03$ (that is between 1\% and 3\% of the size of the computational domain) the largest relative error was 3\%. 



Simulation results demonstrate good computational fidelity of the proposed closure. Therefore, the exact dynamics of $Q$ (Eq. (\ref{eq:Q-exact})) can be well reproduced by the coarse-scale model given by Eq. (\ref{kinetic-final}). The accuracy can be further improved by using a more sophisticated deconvolution closure \cite{PBG, PBC, BP12}. Error estimates for this closure are available \cite{BP12}. As noted above, the Fourier transform of $Q$ with respect to the velocity-conjugated variable $\bz$ can be interpreted as a spatially averaged one-particle distribution function. The coarse-scale dynamics of this function, obtained by taking the Fourier transform of Eq. (\ref{kinetic-final}), is also reasonably accurate.  However,  it is not clear whether this dynamics is dissipative. It is possible that spatial averaging alone is insufficient for constructing a kinetic model that increases a suitable entropy functional. In this regard, we note that all spatial averages depend on the initial positions $\bq_j^0$ and momenta $\bp_j^0$, since $\bq_j(t)=\bq_j(t, \bq_1^0, \bq_2^0, \ldots, \bq_n^0,
\bp_1^0, \ldots, \bp_N^0)$, and similarly for momenta. Choosing a probability density $f(\bq_1^0, \bq_2^0, \ldots, \bq_n^0,
\bp_1^0, \ldots, \bp_N^0)$ one can define ensemble average of $Q$, and then use projection operator method to separate the dissipative component of the ensemble-averaged dynamics, following for example \cite{mori72}. We do not attempt such and extension here. The purpose of this work is to show how spatial averaging can be used to develop a closed-form kinetic theory for single-realization molecular dynamics. The closure construction does not really on the assumption of collision-dominated dynamics. Instead, we use spatial interpolation and related integral approximations. The accuracy of these approximations increases with increasing particle density.  Therefore the resulting kinetic equation should be suitable for dense media. 


\bibliography{main-ref}

\end{document}